\documentstyle[prl,aps,epsfig,multicol]{revtex}
\begin{document}
\draft
\widetext

\newcommand{\be}{\begin{equation}}
\newcommand{\ee}{\end{equation}}
\newcommand{\ber}{\begin{eqnarray}}
\newcommand{\eer}{\end{eqnarray}}

\title{Phase Diagram of Congested Traffic Flow: An Empirical Study}

\author{H. Y. Lee$^{1}$, H.-W. Lee$^{2}$, and D. Kim$^{3}$}
\address{
$^{1}$Department of Physics, Massachusetts Institute of Technology,
Cambridge, MA 02139 \\ 
$^{2}$School of Physics, Korea Institute for Advanced Study,
207-43 Cheongryangri-dong, Dongdaemun-gu, Seoul 130-012, Korea \\
$^{3}$School of Physics, Seoul National University, 
Seoul 151-742, Korea \\
}

\maketitle

\begin{abstract}
We analyze traffic data from a highway section containing
one effective on-ramp.  Based on two criteria, local velocity
variation patterns and expansion (or nonexpansion) of
congested regions, three distinct congested traffic states
are identified.
These states appear at different levels of the upstream
flux and the on-ramp flux, thereby generating a phase digram of the
congested traffic flow.
Observed traffic states are compared with recent 
theoretical analyses and both agreeing and disagreeing features
are found.
\end{abstract}
\widetext

\begin{multicols}{2}

\narrowtext

In the last decade, there has been growing interest in the traffic flow,
which is partly motivated by the fact that the traffic flow is
an easily perceivable realization of heavily studied
driven nonequilibrium systems~\cite{Shmittmann95Book}. 
Another important motivation
is the hope that complex traffic behaviors may be
understood with the help of physical approaches.
Such understanding can be used to optimize traffic and
even to  forecast traffic situations.

A reasonable first step along this line of thinking will be
the classification of distinct traffic states and
separate investigation of their properties.
Various traffic models are proposed~\cite{Nagel96PRE} and 
compared with real traffic data~\cite{Kerner96PRE1}. 
With the help of these models,  
free flow and so called wide traffic jams are well understood.
On the other hand, nature of congested traffic flow (or
synchronized flow), which appears near road inhomogeneities mostly,
yet remains unclear despite various 
empirical~\cite{Kerner96PRE2,Kerner97PRL,Kerner98PRL} and
theoretical~\cite{Lee98PRL,Helbing98PRL,Helbing99PRL,Lee99PRE,%
Tomer00PRL} efforts.

A recent theoretical study~\cite{Helbing99PRL} proposed
an intriguing possibility that the congested traffic flow
may not be a single dynamic phase but rather a collection
of multiple phases, each of which is realized under different
conditions. Similar conclusion is also reported
from the investigation of another theoretical model~\cite{Lee99PRE}.
In the empirical investigations, however,
although qualitatively distinct congested traffic states are 
reported~\cite{Kerner96PRE2,Treiber00preprint}, 
no empirical evidence is
found for the existence of any characteristic parameters 
that distinguish their appearance conditions~\cite{Kerner98PRL}.

\begin{figure}
\begin{center}
\epsfig{file=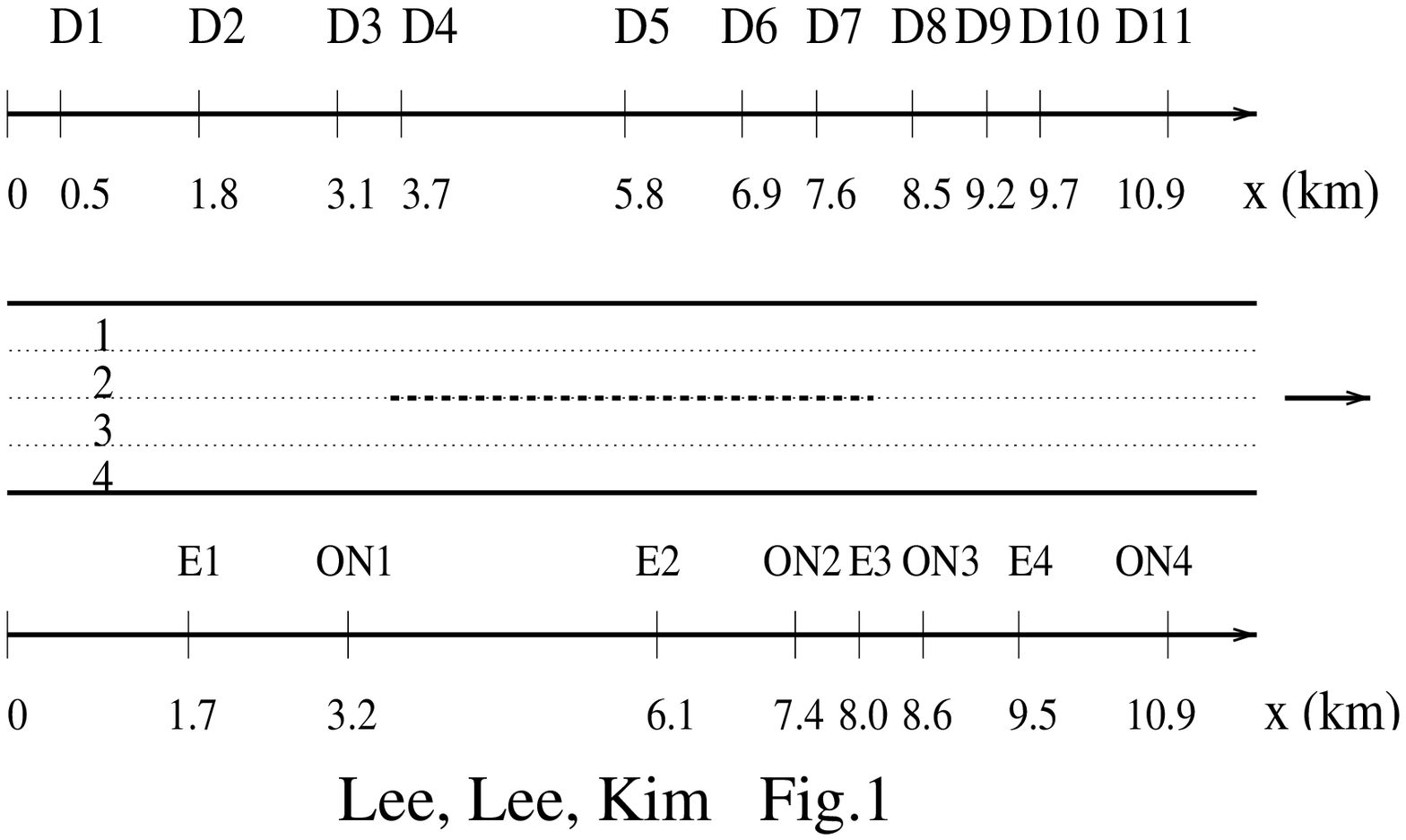,clip=,width=0.95\columnwidth}
\end{center}
\caption[]{
Schematic diagram of a road section in the Olympic Highway in Seoul.
Locations of detectors (D$n$), on-ramps (ON$n$), and
off-ramps (E$n$) are marked. The dashed line 
in the middle denotes the lane divider and the arrow
indicates the driving direction.   
}
\label{fig:geometry}
\end{figure}

In this paper, we report empirical investigation of
traffic congestion in a highway section containing
one effective on-ramp. 
Details of the section are given in our preliminary
report~\cite{Lee99preprint}, so we provide here only
a brief description (Fig.~\ref{fig:geometry}).
All ramps are connected to the outermost lane (lane 4) and 
a stretch of lane divider (from $x=3.5$ to 8.3 km, 
dashed line in Fig.~\ref{fig:geometry}) blocks 
lane change from the two outer lanes (lane 3 and 4) 
to the two inner ones (lane 1 and 2) and vice versa.
In  a short road segment near the end of the lane divider
(from $x=8.3$ km to the location of the detector D9 approximately),
many vehicles in the outer lanes switch
into the inner lanes, which is also enhanced by vehicle flux
through the on-ramp ON3 at $x=8.6$ km.
As a result, this segment works
as an {\it effective} on-ramp region for the traffic flow
of the inner lanes and traffic congestion often occurs
in the road section with the lane divider,
where the inner lanes are decoupled from the outer ones.

\begin{figure}
\begin{center}
\epsfig{file=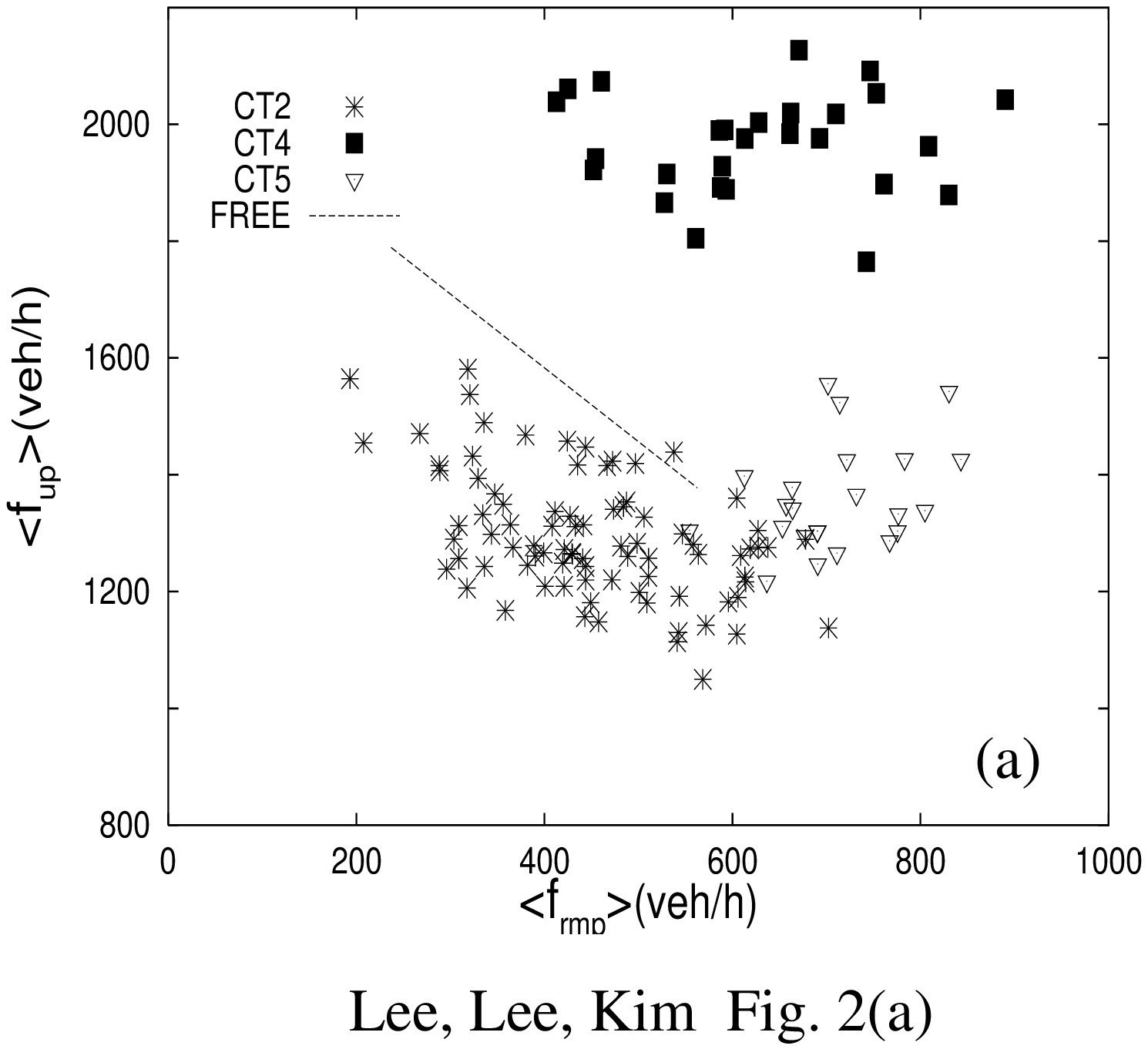,clip=,width=0.95\columnwidth}
\epsfig{file=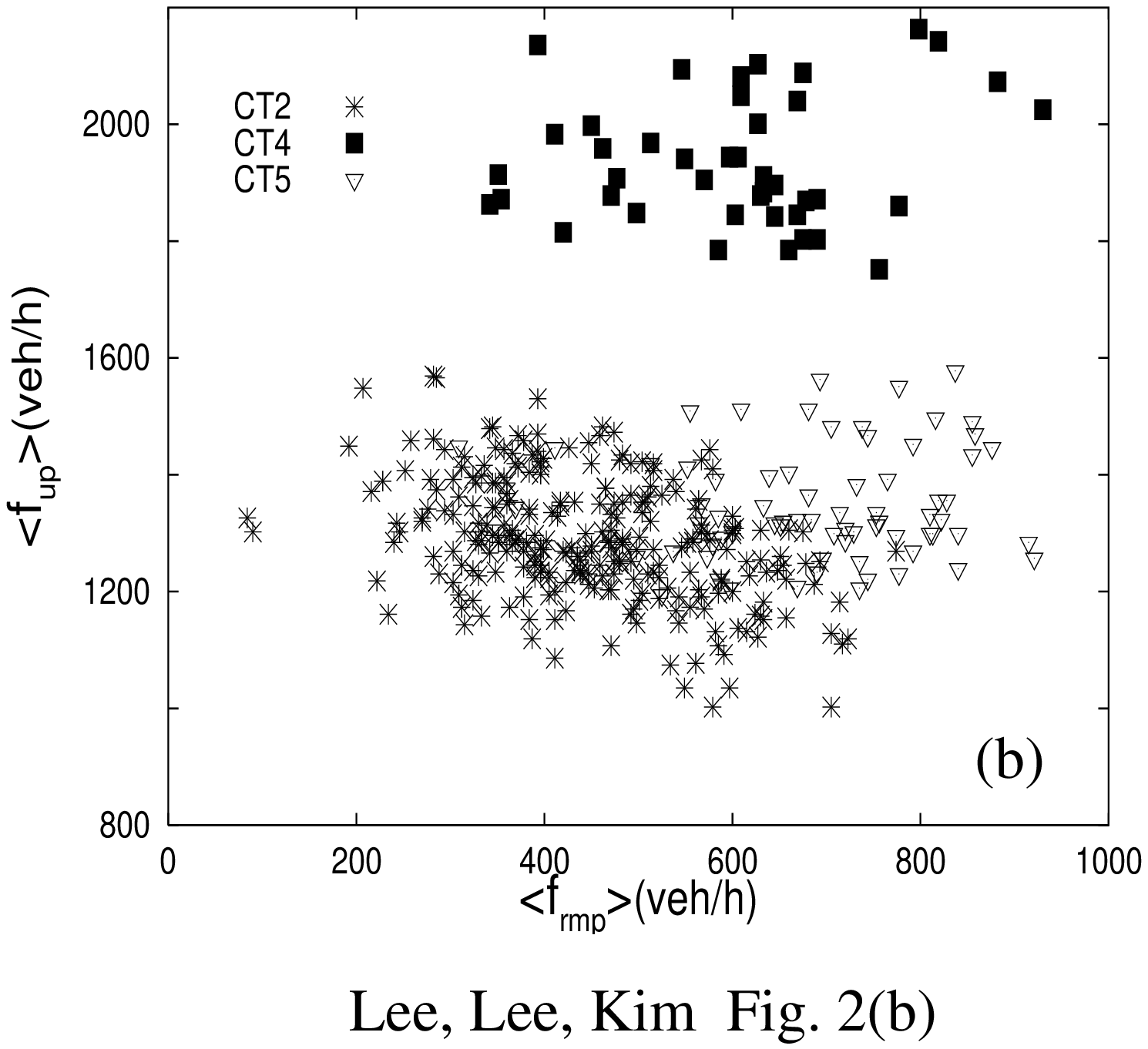,clip=,width=0.95\columnwidth}
\end{center}
\caption[]{
(a) Empirical phase diagram of the congested traffic flow.
$\langle f_{\rm up} \rangle$ and $\langle f_{\rm rmp} \rangle$
represent the average upstream and on-ramp flux values
over the time interval during which a particular 
congested traffic state is maintained (also lane averaged).
The dashed line is an empirical estimation of 
the free flow phase boundary below which 
the free flow can remain linearly stable. 
(b) Same diagram using 10 min averaged flux values.
}
\label{fig:phasediagram}
\end{figure}

We investigate the traffic congestion in the inner lanes 
using the 30 second averaged traffic data from June to September, 1999
(a total of 107 days, a much larger data set
compared to 14 days in Ref.~\cite{Lee99preprint}).
All quantities below are averaged over the two inner lanes.
For each realization of a particular congested traffic state,
which is stably maintained about 30 min or longer,
the effective ramp flux  $f_{\rm rmp}$ [defined as
the difference between two flux values measured at
the detectors D10 and D7, $q($D10$)-q($D7$)$]
and the upstream flux $f_{\rm up}$ 
[$q($D$n)$ is used when the congestion
extends up to D$n+1$] are averaged over the time
interval of its duration and 
the resulting average values $\langle f_{\rm rmp}\rangle$,
$\langle f_{\rm up}\rangle$ are marked in Fig.~\ref{fig:phasediagram}(a).
Note that three congested traffic states, 
which we call CT2, CT4, and CT5 states, respectively (see below),
occupy distinct regions, providing a supporting evidence for
the prediction~\cite{Helbing99PRL,Lee99PRE} that
$f_{\rm up}$ and $f_{\rm rmp}$ are characteristic
parameters of congested traffic flow.
Thus Fig.~\ref{fig:phasediagram}(a) becomes an empirical
phase diagram of the congested traffic flow.
Fig.~\ref{fig:phasediagram}(b) shows an alternative
phase diagram obtained from 10 min averaging of
the flux values instead. Two phase diagrams are qualitatively
the same.

Three congested traffic states, CT2, CT4, and CT5, are
classified according to the two criteria given below.
It is previously reported~\cite{Kerner98PRL} that
the congested region may consist of backward (towards upstream)
traveling clusters and the size of the clusters grow
spontaneously during their backward propagation.
As a result, large amplitude oscillation of velocity develops
spontaneously. 
On the other hand, recent theories~\cite{Helbing99PRL,Lee99PRE}
predict that large velocity oscillation may or
may not develop. Thus our criterion (i) is
whether
such spontaneous growth of velocity oscillation appears (CT5)
or not (CT2, CT4). 
This criterion can be examined by comparing
temporal variation of velocity at different detectors. 

In the same theoretical works, both expanding and nonexpanding
traffic states are predicted, and our criterion (ii)
is whether the congested region expands monotonically (CT4, CT5)
or not (CT2). Mathematically the expansion rate of 
the congested region is proportional to the degree
of flux mismatch $f_{\rm up}+f_{\rm rmp}-f_{\rm down}$,
where $f_{\rm down}$ measures the outflow from
the congested region [$q($D10$)$ is used].
Thus the comparison of 
$\langle f_{\rm up}\rangle+\langle f_{\rm rmp}\rangle$
and $\langle f_{\rm down} \rangle$ can be used as a more
objective application of the criterion (ii).

In all three states, the density-flow relations show
fluctuating behavior (such as Fig.~2(c) in Ref.~\cite{Kerner96PRE2})
and the velocity variations in the lane 1 and 2 are
synchronized (such as Fig.~2 in Ref.~\cite{Kerner97PRL}).
Since these properties are already reported in other publications,
we do not present similar figures here.    
Below we discuss other properties of the congested traffic states. 

\begin{figure}
\begin{center}
\epsfig{file=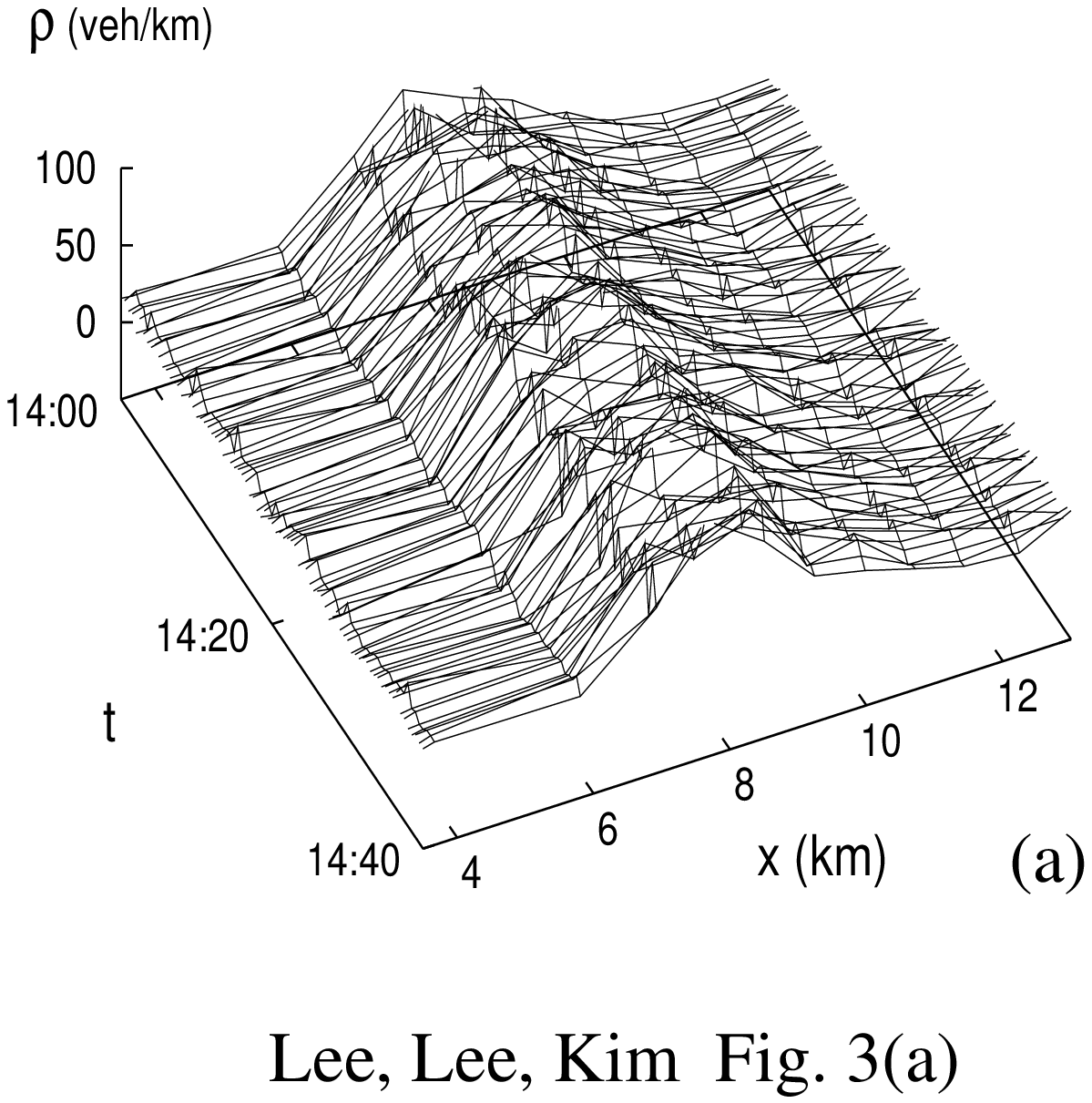,clip=,width=0.95\columnwidth}
\epsfig{file=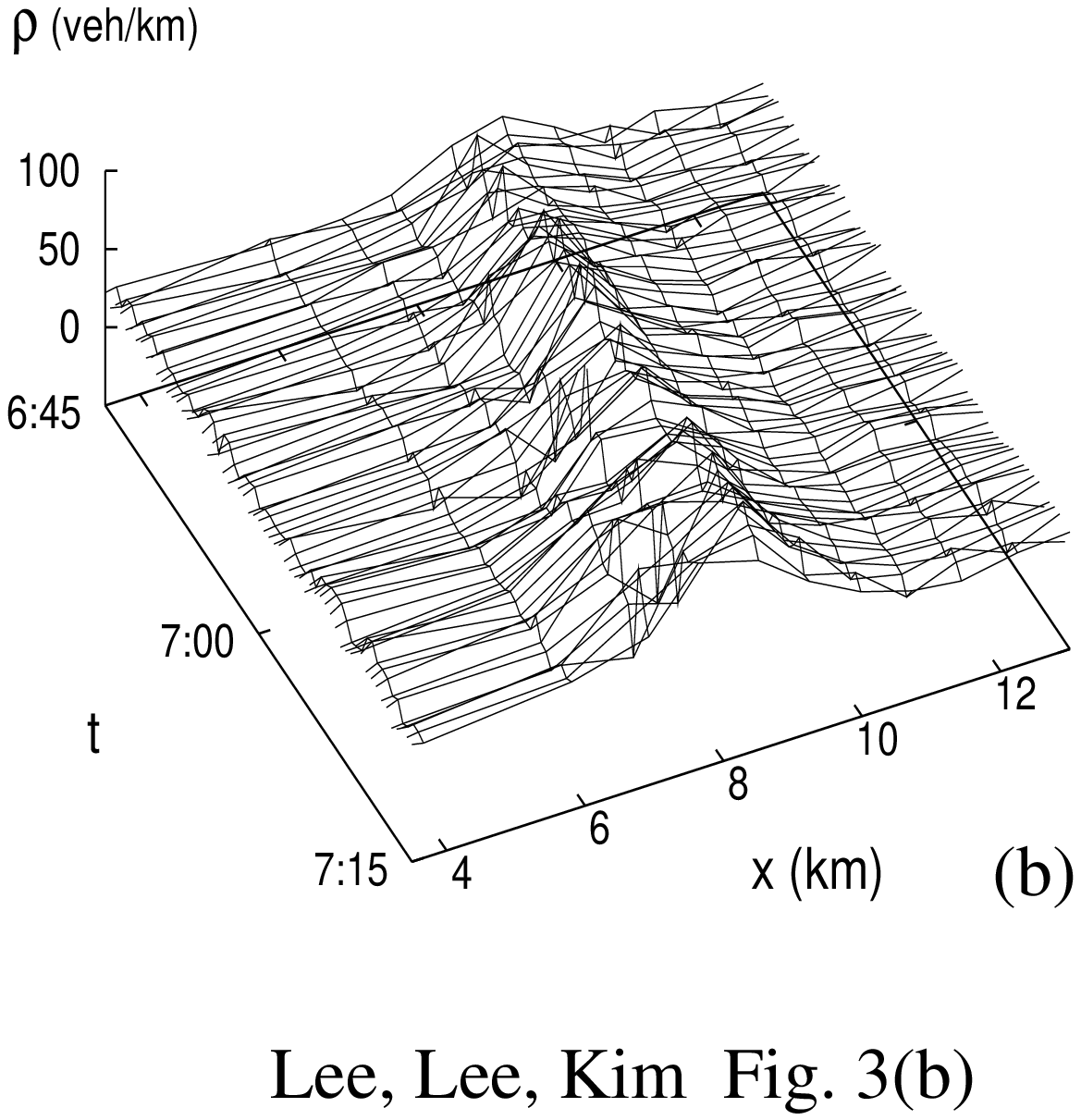,clip=,width=0.95\columnwidth}
\epsfig{file=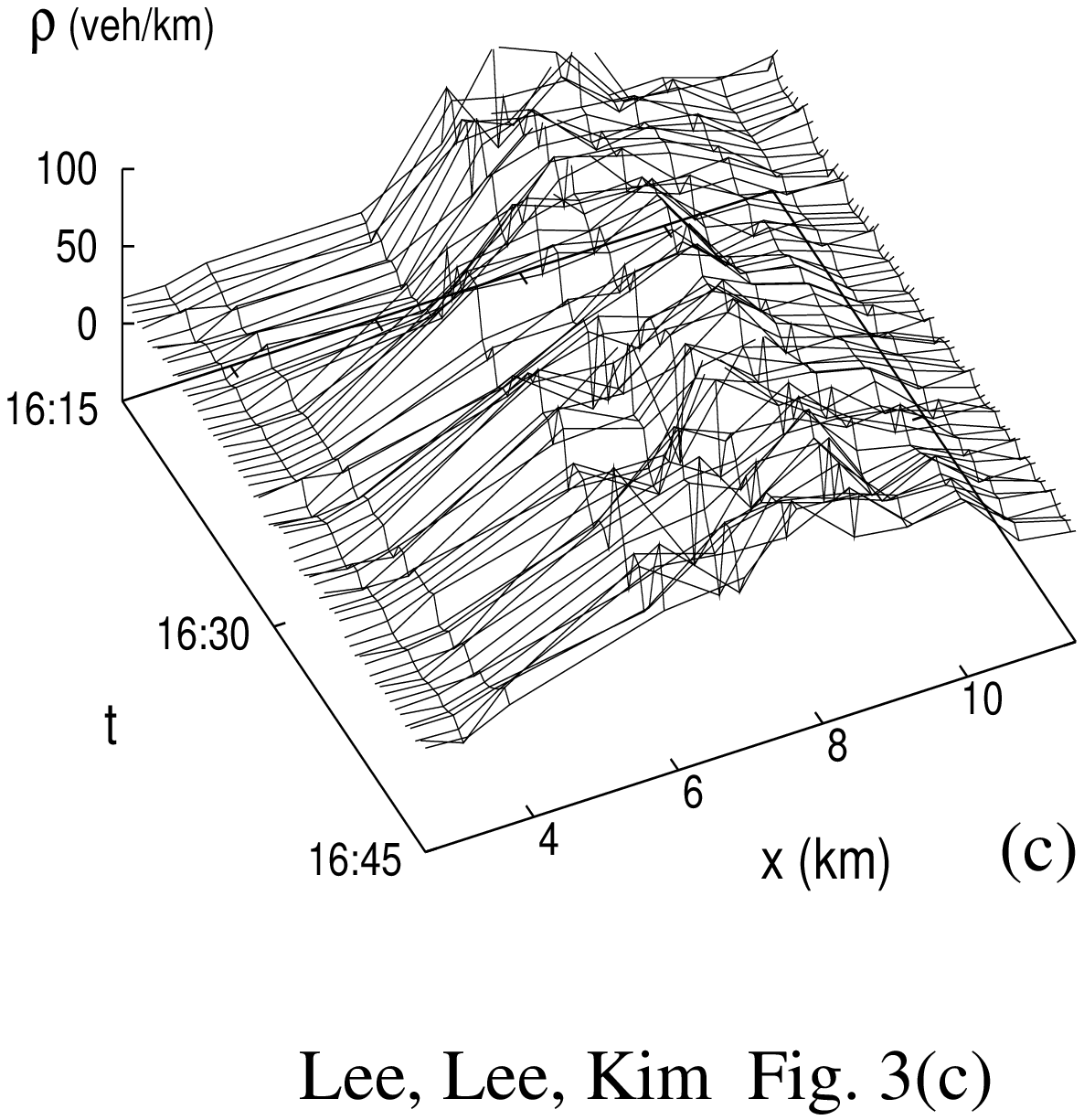,clip=,width=0.95\columnwidth}
\end{center}
\caption[]{
The 3d density profile of the CT2 state (a), CT4 state (b),
and CT5 state (c).
}
\label{fig:3dplot}
\end{figure}

Fig.~\ref{fig:3dplot}(a) depicts the density profile $\rho(x,t)$ of
the CT2 state and its evolution with time,
where $\rho(x,t)$ is evaluated by $q(x,t)/u(x,t)$ 
and $u(x,t)$ is the harmonic mean velocity~\cite{commentU} 
over 30~s intervals.
The spontaneous growth of the velocity oscillation
mentioned in the criterion (i)
does not appear. The flux mismatch 
$\langle f_{\rm rmp}\rangle+\langle f_{\rm up}\rangle
-\langle f_{\rm down}\rangle$ is negligible and  
the congested region does not expand.
Empirical data also show that the length of the congested region 
increases with $\langle f_{\rm rmp}\rangle$ and its dependence
on $\langle f_{\rm up}\rangle$ is rather weak~\cite{commentCT2oscillation}.

Fig.~\ref{fig:3dplot}(b) shows the CT4 state. 
The spontaneous growth of the velocity oscillation does not appear.
Due to the large flux mismatch (typical value of
$\langle f_{\rm up}\rangle+\langle f_{\rm rmp}\rangle
-\langle f_{\rm down}\rangle$ is 500-600 veh/h),
the congested region expands with time. The expansion
rate ranges from 3 to 9 km/h and
increases with increasing flux mismatch.
An interesting property is that the outflow 
$\langle f_{\rm down}\rangle$ is practically independent 
of $\langle f_{\rm up} \rangle$ and $\langle f_{\rm up} \rangle$,
and remains almost universal.
The average of $\langle f_{\rm down}\rangle$ over 
the 28 realizations of the CT4 state 
in Fig.~\ref{fig:phasediagram}(a) 
is 2010 veh/h and its standard deviation is about 65 veh/h, which
is much smaller than the spread of $\langle f_{\rm up} \rangle$ and 
$\langle f_{\rm rmp} \rangle$.
To our knowledge, this is the first empirical indication of
the universal outflow from the congested flow 
near an on-ramp~\cite{commentUniversalOutflow}.

\begin{figure}
\begin{center}
\epsfig{file=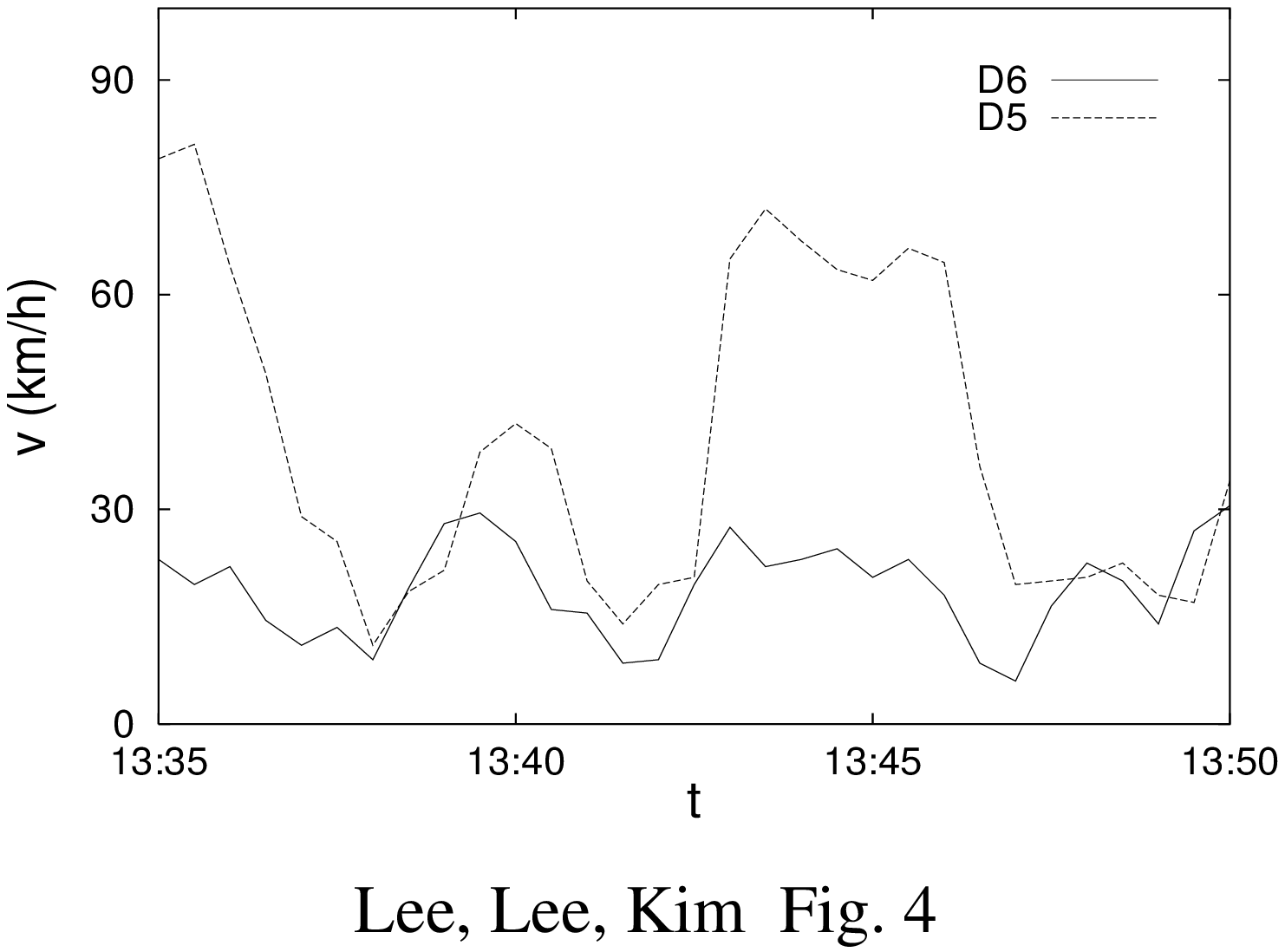,clip=,width=0.95\columnwidth}
\end{center}
\caption[]{
Spontaneous growth of the velocity oscillation
inside the congested region of the CT5 state. 
The velocity evolution at D6 is
shifted to the right by 5 min for clear comparison.
}
\label{fig:spontaneousgrowth}
\end{figure}

Fig.~\ref{fig:3dplot}(c) portrays the CT5 state.
An important feature of the CT5 state is
the spontaneous growth of the velocity oscillation
inside the congested region.
Fig.~\ref{fig:spontaneousgrowth} shows the temporal variation
of the velocity at D5 and D6. The graph for D6 is shifted
to the right by 5 min for comparison. 
Note that the velocity evolutions
at the two detectors are correlated after the time shift
for D6, and that the velocity oscillates with
larger amplitude at D5.
These features imply that velocity wave propagates backward
(towards upstream) and its amplitude grows during
its propagation. 

Regarding the criterion (ii), the congested region
expands as shown in Fig.~\ref{fig:3dplot}(c). 
The upstream front of the congested region initially
locates between D5 and D6 and later between D4 and D5. 
We mention that the flux at D4 remains quasi-stationary
during the depicted time interval. Thus the expansion
is not due to the increase of $f_{\rm up}$ but
due to the flux mismatch. 
Compared to the CT4 state, however, it turns out that
the flux mismatch (typically 200-250 veh/h) is considerably
smaller, which implies a slower expansion 
of the congested region. 
We estimate the expansion rate by the flux mismatch
divided by the density difference at the upstream front
of the congested region, and find it
ranges 2-4 km/h~\cite{commentSlowExpansion}.
The outflow of this state is not universal.

It is interesting to compare empirically identified states 
with theoretically predicted 
states~\cite{Lee98PRL,Helbing98PRL,Helbing99PRL,Lee99PRE}.
Application of the criteria (i,ii) to both
empirical and theoretical states leads to
the following pairing between empirical and theoretical states: 
the CT2 state with the theoretically predicted pinned localized cluster (PLC)
state, the CT4 state with the homogeneous congested
traffic (HCT) state, and the CT5 state with
the oscillating congested traffic (OCT) state.

The pairing motivates further comparison between
the paired states.
For the CT4-HCT pair, we note that
the universal outflow is predicted for the HCT state~\cite{Helbing99PRL} 
and the same property is observed for the CT4 state,
which strongly motivates the identification of
the CT4 state with the HCT states.
For the CT5-OCT pair, on other hand,
while the outflow is universal for the OCT state~\cite{Helbing99PRL},
it is not for the CT5 state.
Thus properties of these two states are only
in partial agreement.
And for the CT2-PLC pair, we find one quantitative
difference: the congested region of the CT2 state is
considerably wider than that of the PLC state in
Refs.~\cite{Helbing99PRL,Lee99PRE}. 
The resolution of this discrepancy 
may require improved traffic theories.

We also compare the theoretical and empirical
phase diagrams. 
It is predicted in Ref.~\cite{Helbing99PRL} 
that when the upstream flux and
the ramp flux are maintained at {\it strictly} constant values,
$f_{\rm up}$ and $f_{\rm rmp}$, respectively,
the phase boundary between the flux matching
and flux mismatching states is practically identical
to the stability boundary of the free flow,
below which the free flow can remain {\it linearly} stable,
and given by the line $f_{\rm up}+\alpha f_{\rm rmp}=\tilde{Q}_{\rm out}$,
where $\alpha=1$ and $\tilde{Q}_{\rm out}$ is a constant
whose value is almost identical to the universal outflow
of the HCT state
($\alpha$ is predicted to be a little larger 
than 1 in Ref.~\cite{Lee99PRE}).

Empirical determination of the free flow stability boundary
is not an easy task since the free flow near the boundary
is quite vulnerable to fluctuations.
In the empirical phase diagram [Fig.~\ref{fig:phasediagram}(a)],
the dashed line 
($\langle f_{\rm up} \rangle+\alpha \langle f_{\rm rmp} \rangle
={\tilde Q}_{\rm out}$, where $\alpha\approx 1.3$ and
${\tilde Q}_{\rm out}\approx 2100$ veh/h)
is an empirical {\it estimation} of
the stability boundary.
Here the values of $\alpha$ and ${\tilde Q}_{\rm out}$ are
reliable up to their first significant digits
and their second significant digits are
rather uncertain. 
Within this accuracy, $\alpha$ is approximately one and
the value of ${\tilde Q}_{\rm out}$ is 
close to the universal outflow of the CT4 state.
It is also pleasing to note that this line divides 
(except for a small number of data points) 
the flux matching and mismatching states,
in agreement with the theoretical prediction~\cite{Helbing99PRL}.
This feature is robust and independent
of details of the boundary estimation method
although the values of $\alpha$
and $\tilde{Q}_{\rm out}$ depend on the details.
We also note that in a later, more refined 
theory~\protect\cite{Treiber00preprint}, 
it is predicted that the PLC and OCT states overlap weakly 
in the phase diagram. Recalling the pairing CT2-PLC and CT5-OCT,
the weak overlap of the CT2 and CT5 states 
in Fig.~\protect\ref{fig:phasediagram}(a)
is in agreement with this prediction. Also the locations of
the empirical and theoretical overlap regions 
in the phase diagram are similar. 

Regarding the phase boundary between the two expanding
traffic states (CT4,CT5), while the recent theory~\cite{Helbing99PRL}
predicts the line $f_{\rm rmp}=const.$ as the phase
boundary between the HCT and OCT states,
Fig.~\ref{fig:phasediagram}(a) shows that
the boundary between the CT4 and CT5 states is
better described by $\langle f_{\rm up}\rangle=const.$
Thus as for the phase boundary between the expanding states,
empirical results and the theoretical prediction
do not agree.

We next discuss an implicit but important conceptual
implication of the recent 
theories~\cite{Lee98PRL,Helbing98PRL,Helbing99PRL,Lee99PRE},
where traffic phases are identified with resulting
final states that traffic flow dynamics lead to
after sufficient transient time.
In the language of nonlinear dynamics, 
all these states correspond to 
stable attractors~\cite{Jackson89Book}  of traffic flow dynamics.
An implication of this concept is
that for a given external condition 
(such as $f_{\rm rmp},f_{\rm up}$ and ramp geometry), 
the resulting traffic state is 
independent of details of the initial traffic state
or its ``evolution history''~\cite{commentAttractor}.
 
\begin{figure}
\begin{center}
\epsfig{file=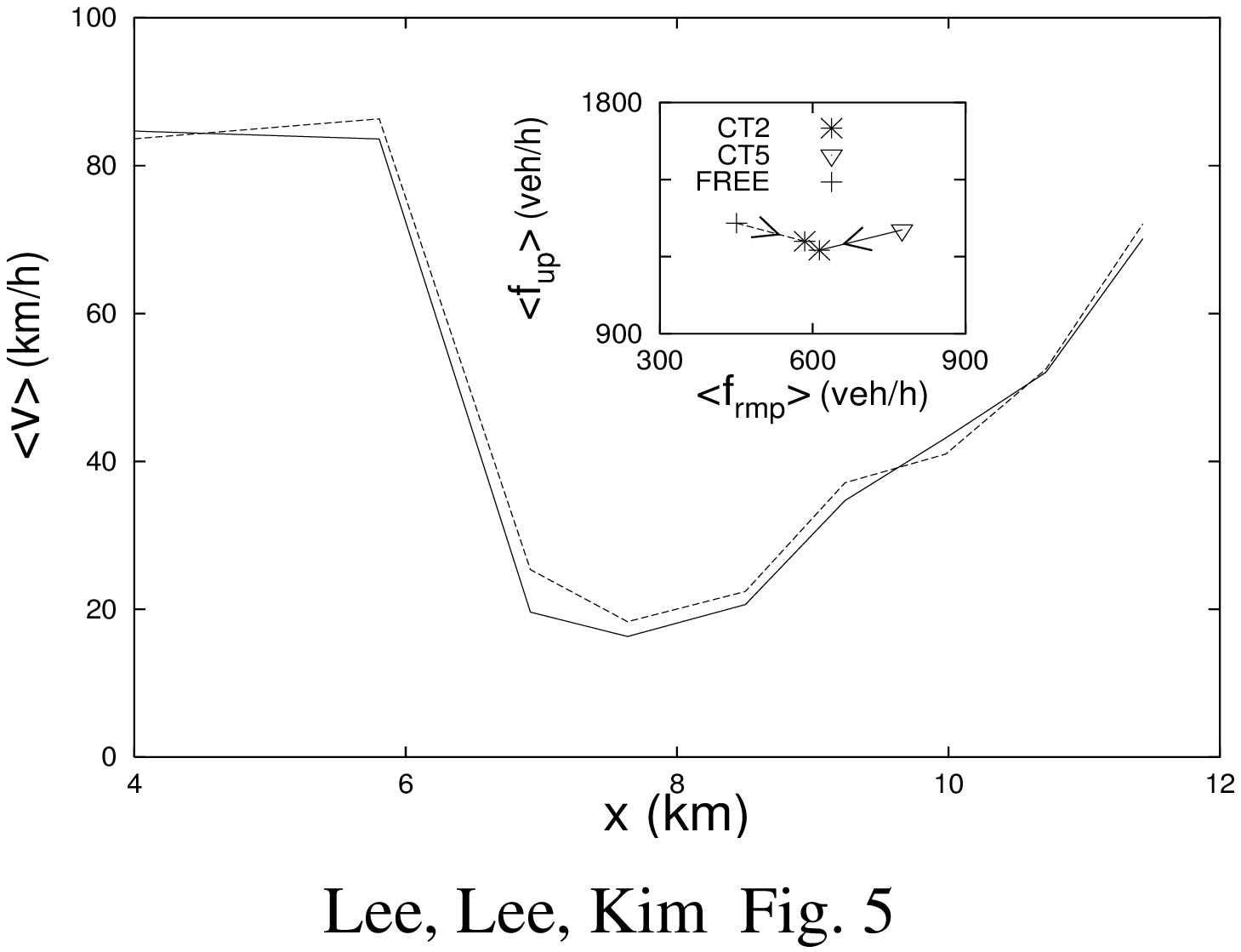,clip=,width=0.95\columnwidth}
\end{center}
\caption{Time averaged velocity profiles of two realizations 
of the CT2 state with almost identical $\langle f_{\rm up}\rangle$
and $\langle f_{\rm rmp}\rangle$ but with
different evolution histories
(solid line for the realization 1 and dashed line for
the realization 2).
Inset: While the realization 1 has developed from the CT5 state 
via a phase transition, the realization 2 has evolved
from the free flow.
}
\label{fig:differentevolution}
\end{figure}

This idea can be tested empirically, for example, by
comparing realizations with the same external conditions
but with different evolution histories. 
Fig.~\ref{fig:differentevolution} compares the (time averaged)
velocity profile of two realizations of
CT2 state, both with almost the same
$\langle f_{\rm rmp}\rangle$ and $\langle f_{\rm up} \rangle$.
But their evolution histories are different:
one has evolved from the free flow and the other
from the CT5 state (inset in Fig.~\ref{fig:differentevolution}).
Note that two profiles almost overlap with each other
despite qualitatively different evolution histories.
This insensitivity to
the evolution history confirms the conceptual implication
of the recent theoretical works.

Obviously there is a fundamental difference between
theoretical and empirical situations:
while theories~\cite{Lee98PRL,Helbing98PRL,Helbing99PRL,Lee99PRE} 
assume strictly constant upstream and ramp fluxes, 
in real traffic the fluxes fluctuate all the time.
Thus in order to make a rigorous comparison with theories,
a good understanding of fluctuation effects is necessary.
However such an understanding is not available at present, so
in this paper, the fluctuation
effects are essentially ignored and the time
averaged values $\langle f_{\rm rmp}\rangle$ and
$\langle f_{\rm up}\rangle$ are instead used in the analysis.
Hence in a strict sense, the present investigation
is only a correlation analysis between the time averaged
flux values and the traffic states that are maintained
for a sufficiently long time.
In retrospect, however, it turns out that
many empirical results can be explained by the theories 
if one accepts that $\langle f_{\rm rmp}\rangle$ 
and $\langle f_{\rm up}\rangle$ play the roles of 
the constant ramp flux and upstream flux, respectively,
assumed in the theories.
Hence it appears that the neglect of the fluctuation effects
can be justified at least {\it a posteriori}.

Its justification can be partly understood from
two arguments: firstly, while the fluctuation effects are usually
crucial near the phase boundary, traffic states far from 
the phase boundary are relatively insensitive to fluctuations.
Thus when a traffic state occupies a sufficiently
wide region in the phase diagram, the use of
the time averaged flux values for the phase diagram
can be justified.
Secondly, dominant contributions
to flux fluctuations come from short time scale of order 1 min,
and in the time scale of about ten min or longer, 
flux variation is almost quasi-stationary.
Thus provided that the relevant time scale of
a traffic state (not very close to the phase boundary)
is longer than 1 min, the short time scale fluctuations
will be effectively averaged out by the traffic dynamics
itself and can be neglected indeed.

In support of these arguments, we find that 
our results are not sensitive 
to the length of the averaging time interval 
as long as it is sufficiently longer than 1 min.
When the same analysis is repeated with 10 min averaging,
Fig.~\ref{fig:phasediagram}(b) is obtained. 
Note that this new phase diagram is
qualitatively the same as the former one [Fig.~\ref{fig:phasediagram}(a)]
and all discussions above remain unchanged.
Changes occur only in a quantitative level.
With the 10 min averaging, the standard deviation of
the outflow $\langle f_{\rm down} \rangle$ from
the CT4 state increases to about 85 veh/h.

We remark on a few details of the analysis. Firstly, there exists
another on-ramp (ON4) at 2.3 km downstream from ON3. 
Sometimes vehicle flux through this on-ramp causes 
traffic congestion and the resulting congestion extends 
to the region studied in this paper. 
All such events are excluded from 
the present analysis to focus on one particular
inhomogeneity~\cite{commentCT3}. 
Secondly, in the analyzed road section, there are two spots
(one between D5 and D6 and the other near D7),
where the lane divider is imperfect.
All time intervals with non-negligible vehicle fluxes
through those spots are not included 
in the analysis.

In summary, three congested traffic states are identified
based on local velocity variation patterns
and expansion (or nonexpansion) of the congested region. It is found
that the appearance of these congested traffic states
is strongly correlated with the time averaged flux values,
$\langle f_{\rm up} \rangle$ and $\langle f_{\rm rmp}\rangle$,
providing a strong supporting evidence to 
the prediction~\cite{Helbing99PRL,Lee99PRE} that these flux values
are characteristic parameters of the congested traffic flow.
An empirical phase diagram is constructed and compared with
theoretical predictions. The prediction on
the phase boundary between the flux matching and mismatching states
is consistent with the empirical phase diagram and 
the prediction of the universal outflow are confirmed.
However some deviations from theoretical predictions
are also found.
Lastly we mention that there exist regions in the 
$\langle f_{\rm rmp}\rangle$-$\langle f_{\rm up}\rangle$ plane
which are not probed. Thus it is possible that
additional congested traffic states exist 
in those regions. 

We thank Young-Ihn Lee and Seung Jin Lee for providing 
the traffic data, Sung Yong Park for fruitful discussions.
This work was supported in part by the BK21 Project of
Korean Ministry of Education.

\end{multicols}


\begin{thebibliography}{7}
\bibitem{Shmittmann95Book} B. Shmittmann and R. K. Zia, in 
  {\it Phase Transitions and Critical Phenomena}, Vol. 17, 
  edited by C. Domb and J. Lebowitz (Academic, New York, 1995).
\bibitem{Nagel96PRE} K. Nagel and M. Schreckenberg, J. Phys. I (France)
  {\bf 2}, 2221 (1992); B. S. Kerner and P. Konh\"{a}user, Phys. Rev. E 
  {\bf 48}, R2335 (1993); M. Bando, K. Hasebe, A. Nakayama, A. Shibata,
  and Y. Sugiyama, {\it ibid}. {\bf 51}, 
  1035 (1995); S. Krauss, P. Wagner, and C. Gawron, {\it ibid}. 
  {\bf 55}, 5597 (1997).
\bibitem{Kerner96PRE1} B. S. Kerner and H. Rehborn, Phys. Rev. E {\bf 53},
  R1297 (1996), and references cited therein.
\bibitem{Kerner96PRE2} B. S. Kerner and H. Rehborn, Phys. Rev. E {\bf 53},
  R4275 (1996).
\bibitem{Kerner97PRL} B. S. Kerner and H. Rehborn, Phys. Rev. Lett.
  {\bf 79}, 4030 (1997).
\bibitem{Kerner98PRL} B. S. Kerner, Phys. Rev. Lett. {\bf 81}, 3797 (1998). 
\bibitem{Lee98PRL} H. Y. Lee, H.-W. Lee, and D. Kim, Phys. Rev. Lett.
  {\bf 81}, 1130 (1998).
\bibitem{Helbing98PRL} D. Helbing and M. Treiber,
  Phys. Rev. Lett. {\bf 81}, 3042 (1998).
\bibitem{Helbing99PRL} D. Helbing, A. Hennecke, and M. Treiber,
  Phys. Rev. Lett. {\bf 82}, 4360 (1999).
\bibitem{Lee99PRE} H. Y. Lee, H.-W. Lee, and D. Kim, Phys. Rev. E {\bf 59},
  5101 (1999).
\bibitem{Tomer00PRL} E. Tomer, L. Safonov, and S. Havlin,
  Phys. Rev. Lett. {\bf 84}, 382 (2000).
\bibitem{Treiber00preprint} M. Treiber, A. Hennecke, and D. Helbing,
  cond-mat/0002177.
\bibitem{Lee99preprint} H. Y. Lee, H.-W. Lee, and D. Kim,
  unpublished (cond-mat/9905292).
\bibitem{commentU} The harmonic mean velocity $u$
  $[1/u\equiv (1/N) \sum_{j=1}^{N} 1/v_j]$ is used
  instead of $v$ $[v\equiv (1/N)\sum_{j=1}^{N}v_j]$
  since $q/v$ systematically underestimates the density
  when the velocity fluctuations are significant
  (see for example L. Neubert, L. Santen, A. Schadschneider,
  M. Schreckenberg, cond-mat/9905216) 
  whereas $q/u$ takes a better account of the fluctuation
  effects. Also it can be shown that when 
  both density and flux are temporally averaged
  [for example, 
  $\rho(x,t)=(1/T)\int^{T/2}_{-T/2}dt' \sum_j \delta (x_j(t+t')-x)$,
  $q(x,t)=(1/T)\int^{T/2}_{-T/2}dt' \sum_j v_j(t+t')\delta (x_j(t+t')-x)$],
  the ratio $q/\rho$ is exactly equal to $u$. 
\bibitem{commentCT2oscillation} Usually the boundary between
  the congested region and the upstream free region locates
  {\it between} detectors. But due to the dependence of
  the congested region size on $\langle f_{\rm rmp} \rangle$, 
  the boundary will locate {\it at} a particular detector
  when $\langle f_{\rm rmp}\rangle$ is close to a certain
  special value.
  In such a case, small fluctuations of the boundary
  can cause large amplitude fluctuations of velocity
  at that special detector since the velocity varies 
  greatly across the boundary.  
  Such large velocity fluctuations are indeed
  observed in narrow ranges of $\langle f_{\rm rmp} \rangle$
  (for example at D6 when $\langle f_{\rm rmp} \rangle=330\pm 30$
  veh/h). 
  We remark that these fluctuations should be distinguished
  from the velocity oscillations~\protect\cite{Kerner98PRL} 
  in the criterion (i). While the former originates 
  from the boundary fluctuations and is localized near the boundary,
  the latter grows continuously inside the congested region
  and is extended over relatively wide congested region.
\bibitem{commentUniversalOutflow} At 30~s scale,
  $f_{\rm down}$ $[q($D$10)]$ fluctuates within the range
  $2010\pm 140$ veh/h, 
  which is equivalent to $16.8\pm 1.2$ veh/30~s. 
  Here the inaccuracy of $\pm 1.2$ veh/30~s is quite a small number
  considering that the vehicle number counting by detectors
  always results in integer numbers and thus the inaccuracy
  of $\sim \pm 1$ veh/30~s is unavoidable in 30~s data.
\bibitem{commentSlowExpansion} It may be hard to detect
  slow expansion from the 3d density plot due to 
  the finite ($\sim$ 1 km) spatial resolution given
  by the average detector spacing. For this reason,
  this state is incorrectly interpreted as a nonexpanding
  state in our preliminary report~\protect\cite{Lee99preprint}
  (using only 14 day traffic data),
  where a different name, CT1, is used for the same state.
\bibitem{Jackson89Book} E. A. Jackson, {\it Perspectives of
  Nonlinear Dynamics} (Cambridge University Press, Cambridge,
  England, 1989), Vol. 1.
\bibitem{commentAttractor} When there exist multiple (locally)
  stable attractors for a given external condition, initial
  traffic state is not completely irrelevant. But even in this case,
  the initial state affects only the choice of a particular
  attractor as a final state. Or in the language of
  nonlinear dynamics, details of the initial state are irrelevant 
  as long as the initial state remains within 
  the basin of attraction~\protect\cite{Jackson89Book}. 
\bibitem{commentCT3} It is found that the on-ramp flux 
  through ON4 may generate still another type of 
  congested traffic state (called CT3 state in Ref.~\cite{Lee99preprint}),
  which is not included in the present analysis since
  this state appears under different road geometry.


\end{thebibliography}
\end{document}